\documentclass[prb,showpacs,floatfix,twocolumn]{revtex4}
\usepackage{epsfig,epsf}
\newcommand{\R}{{\mathbb{R}}}

\usepackage{amsfonts,amscd,epsfig}

\begin{document}
\title{Radiation emission due to fluxon scattering
on an inhomogeneity in a large two-dimensional Josephson junction }
\author{Ivan O. Starodub}
\email{starodub@bitp.kiev.ua}
\affiliation{
Bogolyubov Institute for Theoretical Physics,
National Academy of Sciences of Ukraine,
vul. Metrologichna 14B, 
03680 Kyiv, Ukraine
}

\author{Yaroslav Zolotaryuk}
\email{yzolo@bitp.kiev.ua}
\affiliation{
Bogolyubov Institute for Theoretical Physics,
National Academy of Sciences of Ukraine,
vul. Metrologichna 14B, 
03680 Kyiv, Ukraine
}
\date{\today}

\begin{abstract}
Interaction of a fluxon in the two-dimensional large Josephson
junction with the finite-area inhomogeneity is studied
within the sine-Gordon theory. 
The spectral density of the emitted plane waves is computed exactly 
for the rectangular and rhombic inhomogeneities. The total emitted
energy as a function of the fluxon velocity exhibits at least one 
local maximum.
Connection to the previously studied limiting cases including the
point impurity and the one-dimensional limit has been performed. An 
important feature of the 
emitted energy as a function of the fluxon velocity is a clear maximum
(or maxima). The dependence of these maxima on the geometric properties
of the impurity has been studied in detail. 
\end{abstract}
\maketitle

\section{Introduction}\label{intro}

Studies of the fluxon (Josephson vortex) dynamics in large Josephson 
junctions\cite{barone82,likharev86} (LJJs) is an important problem in 
modern superconductivity. The LJJs can be spatially inhomogeneous 
either due to the production defects or can be manufactured in such a 
way on purpose. Thus, the problem of the fluxon interaction with the 
spatial inhomogeneity (microshort,microresistor, Abrikosov vortex etc.) 
is of remarkable 
importance \cite{ms78pra,gf84jetp,ag84jetpl,fgb92prb,bs95prb,fg98prb}. 
As a result of the fluxon-impurity interaction the radiation of the 
small-amplitude linear waves (Josephson plasmons) 
occurs\cite{ms78pra,ms79ssc}. The issue of the linear wave radiation 
due to the fluxon collision 
with the spatial inhomogeneity has been studied in detail for the
one-dimensional case (1D). Most of these (both theoretical and 
experimental) studies have focused on the scattering on the point-like inhomogeneity being either a microshort or a  microresistor\cite{ms78pra,kmn88jetp,mstu88jetpl,mu90jap}, 
or a magnetic impurity \cite{kc91prb}. An extended inhomogeneity has 
been investigated\cite{kkc88pla} as well as the interface separating 
two different junctions \cite{kkc87fnt}.
 
An important thing to note is that a 1D Josephson junction is only 
a 1D approximation of the two-dimensional (2D) LJJ of the finite 
width. Thus, a natural question is to take the transverse direction
into account and to study the fluxon scattering on an impurity
in this situation. Moreover, fluxon dynamics in the large area JJ 
is an interesting and important problem in its own right. It has been 
studied in different contexts  such as dynamical properties \cite{loes85jap,lfdfh93prb}, pinning on impurities \cite{sz12pla} 
and applications\cite{gksyn09prb,nk10pc}. However, up to now the 
radiation emission due to the 2D fluxon scattering on the impurity 
has been studied in detail only for the special case of the point-like 
impurity described by the Dirac $\delta$- function\cite{m91pd}. Thus, 
the aim of this paper is to study the properties of the small-amplitude 
wave radiation that appears as a result of the fluxon transmission 
through the inhomogeneity of the general shape.  

The  paper is organized as follows. In the next section, the model 
is described. Section \ref{radiation} is devoted to the studies of 
the radiation emitted due to the fluxon-impurity interaction. 
In the last section, the discussion and conclusions are presented.

\section{The model}\label{model}

We consider fluxon dynamics in the LJJ with spatial inhomogeneities.
The main dynamical variable is the difference between the phases 
$\theta_2(x,y;t)-\theta_1(x,y;t)\doteq \phi(x,y;t)$ of the macroscopic 
wave functions of the the superconducting layers of the junction, 
also known as the Josephson phase. In the bulk of the junction this 
variable satisfies \cite{barone82,likharev86,ms78pra} the equation
\begin{equation}\label{1a}
{\partial_x H_y}-{\partial_y H_x}=j_c [1+f_I(x,y)]\sin \phi+
\frac{\hbar C(x,y)}{2e}{\partial^2_t \phi}~,
\end{equation}
where the function $f_I(x,y)$ describes the critical current change on
the spatial inhomogeneity and the magnetic field components $H_{x,y}$ 
are related to the Josephson phase as
\begin{equation}
H_{x}=-\frac{\hbar}{2e\mu_0 ~l(x,y)}{\partial_y \phi}~,
~~~H_{y}=\frac{\hbar}{2e\mu_0 ~l(x,y)}{\partial_x \phi}~.
\end{equation}
The junction capacitance $C(x,y)$ is spatially inhomogeneous due to the 
impurity. Among other parameters $j_c$ is the critical current density 
away from the impurity,  $e$ is the electron charge, $\mu_0$ is the 
vacuum permeability and $\hbar$ is Planck's constant. The value 
$l(x,y)$ describes the thickness of the layer that allows magnetic field penetration. It varies in space due to the presence of the impurity and
can be written as $l(x,y)=2\lambda_L+d_i(x,y)$, where $\lambda_L$ is
the superconductor London penetration depth and $d_i(x,y)$ is the 
insulating layer thickness. Away from the impurity $d_i(x,y)=d_0=const$
while $d_i(x,y)=d_0+d_1=const$ inside the impurity. For the impurity 
of the general shape that covers a certain segment $\Omega \in \R^2$ 
of the junction one can write
\begin{equation}\label{3b}
f_I(x,y)=\left \{\begin{array}{ccc}
 &\mu_I~~~& \mbox{if}~~ (x,y) \in \Omega~, \\
 &0~~~  &\mbox{if}~~ (x,y) \notin \Omega~.
\end{array} \right .
\end{equation}
Similarly, the spatial change of the magnetic length and capacitance is
given by
\begin{equation}
l(x,y)=\left \{\begin{array}{ccc}
 &d_0+2\lambda_L+d_1~~~& \mbox{if}~~ (x,y) \in \Omega~, \\
 &l_0=d_0+2\lambda_L~~~  &\mbox{if}~~ (x,y) \notin \Omega~.
\end{array} \right .
\end{equation}
and
\begin{eqnarray}\nonumber
&&C(x,y)=C_0\frac{d_0}{d_i(x,y)}=C_0 [1+f_C(x,y)],\\
&&f_C(x,y)=\left \{\begin{array}{ccc}
 &\mu_C=-\frac{d_1}{d_1+d_0}~~~& \mbox{if}~~ (x,y) \in \Omega~, \\
 &0~~~  &\mbox{if}~~ (x,y) \notin \Omega~.
\end{array} \right .
\label{5a}
\end{eqnarray}
where $C_0$ is the junction capacitance per unit area away from the
impurity.
For the sake of convenience the following function can be introduced
\begin{eqnarray}\nonumber
&&{l_0 \over l(x,y)}=1+f_H(x,y)=1+\left \{\begin{array}{cc}
0 &\mbox{if}~~ (x,y) \notin \Omega~\\
\mu_H& \mbox{if}~~ (x,y) \in \Omega~
\end{array} \right. ~,\\
&&\mu_H=\frac{l_0}{l_0+d_1}-1=-\frac{d_1}{d_0+d_1+2\lambda_L}~. \label{6}
\end{eqnarray}
Equation (\ref{1a}) can be rewritten in the dimensionless form by   
normalizing the spatial variables $x$ and $y$ to the Josephson 
penetration depth $\lambda_J$ and the time $t$ to the inverse 
Josephson plasma frequency $\omega_J^{-1}$. As a result, the 
two-dimensional perturbed sine-Gordon (SG) equation is obtained:
\begin{eqnarray}\label{1}
&&\left \{-\partial_x [1+f_H(x,y)] \partial_x-
\partial_y [1+f_H(x,y)] \partial_y +\right . \\
&& \left . +[1+f_C(x,y)]\partial_t^2 \right \}\phi +
\nonumber
[1+f_I(x,y)]\sin \phi=0~.
\end{eqnarray}
For details one might consult the textbooks\cite{barone82,likharev86}. 
The impurity 
is a microshort if $\mu_I>0$, $d_1<0$ and a microresistor if $\mu_I<0$, 
$d_1>0$. Hence $\mu_H/\mu_I>0$ and $\mu_C/\mu_I>0$ both for microshorts 
and microresistors. Taking into account that for the SIS (superconductor-insulator-superconductor) junctions usually \cite{barone82,likharev86} 
the insulating layer thickness $d_0\sim 10 \AA$, while the London 
penetration depth $\lambda_L $ is of the order of several tens of 
$\AA$, the inequality $|\mu_H|<|\mu_C|$ holds.

\section{Radiation emission}\label{radiation}

Fluxon interaction with the spatial inhomogeneity is normally
accompanied with the radiation of the small-amplitude electromagnetic 
waves \cite{ms78pra} (Josephson plasmons). Below we present the general 
scheme for the calculation of the radiation created by the 
fluxon-impurity interaction which is based on the method developed for 
the delta-like impurity \cite{m91pd} or for the respective 1D problems \cite{kmn88jetp,km88pla}. Only the main points of the derivation 
procedure are presented. For the details the interested reader can 
consult the above-mentioned papers.

\subsection{General framework}

Both sides of the SG equation (\ref{1}) can be divided by 
$[1+f_C(x,y)]$, and- as a result it can be rewritten as
\begin{eqnarray}\nonumber
\partial_t^2\phi-\Delta \phi +[1+{\bar f}_{I}(x,y)]\sin \phi = 
{\bar f}_{H}(x,y)\Delta\phi
+\\
+\frac{1}{1+f_C(x,y)}\left [\partial_x f_H(x,y)~ \partial_x \phi +
\partial_y f_H(x,y) ~\partial_y \phi \right ]~,
\label{1b}
\end{eqnarray}
where $\Delta=\partial_x^2+\partial_y^2$ and 
\begin{eqnarray}\nonumber
&&{\bar f}_{I}(x,y)=\frac{1+{f}_{I}(x,y)}{1+{f}_{C}(x,y)}-1=\left 
\{\begin{array}{cc}
0 &\mbox{if}~~ (x,y) \notin \Omega~\\
\bar{\mu}_I& \mbox{if}~~ (x,y) \in \Omega~
\end{array} \right. ,\\
\nonumber
&&{\bar f}_{H}(x,y)=\left \{\begin{array}{cc}
0 &\mbox{if}~~ (x,y) \notin \Omega~\\
\bar{\mu}_H& \mbox{if}~~ (x,y) \in \Omega
\end{array} \right.,~\\
&&{\bar \mu}_{I,H}=\frac{\mu_{I,H}-\mu_C}{1+\mu_C}~.
\end{eqnarray}
We seek the solution of the SG equation (\ref{1}) as a superposition of 
the exact  soliton solution and the plasmon radiation on its background:
$\phi(x,y,t)=\phi_0(x,t)+\psi(x,y,t)$. The spatial inhomogeneity is 
considered as a small ($|\mu_{I,H,C}|\ll 1$) perturbation. Here
$\phi_0(x,t)=4\arctan \left [ \exp \left (\frac{x-vt}{\sqrt{1-v^2}} \right )\right ]$ is
the exact soliton solution of the unperturbed 1D SG equation and
$\psi(x,y,t)$ is the radiative correction, $|\psi| \ll \phi_0$. It is 
convenient to work in the
reference frame that moves with the fluxon velocity $v$:
$\xi=\frac{x-vt}{\sqrt{1-v^2}}$, $\tau=\frac{t-vx}{\sqrt{1-v^2}}$.
In these new variables we have
$\phi_0(x,t)=\phi_0(\xi)=4\arctan \left ( \exp \xi\right )$.

In the moving reference frame the equation that describes the
emitted radiation reads
\begin{equation}\label{3}
\left \{ \partial_{\tau}^2-(\partial_{\xi}^2+\partial_y^2)+\cos 
[\phi_0(\xi)]\right \}\psi =R(\xi,y;\tau)~,
\end{equation}
where the right-hand side of Eq. (\ref{3}) is completely defined by
 the impurity:
\begin{eqnarray}\label{3a}
&&R(\xi,y;\tau)={2} \left [ \left (1-{1 \over 1-v^2}
\frac{{\bar \mu}_H}{{\bar \mu}_I} \right ) {\tanh \xi \over \cosh \xi}
\times ~\right.\\
\nonumber
&&\left .\times {\bar f}_I\left (\frac{\xi+v\tau}{\sqrt{1-v^2}},y\right )
+
{h_H\left (\frac{\xi+v\tau}{\sqrt{1-v^2}},
y\right ) \over \sqrt{1-v^2}~\cosh \xi}   \right ],\\
&& \nonumber
h_H(x,y)=\partial_x f_H(x,y)~.
\end{eqnarray}
In this expression it has been taken into account that
$\sin [\phi_0(\xi)]=\partial_\xi^2\phi_0(\xi)=-2\tanh \xi/\cosh{\xi}$. 
Also, for any two functions of the type $f_\alpha(x,y)$ or 
${\bar f}_\alpha(x,y)$ ($\alpha=I,C,H$) the following equality is true:
$f_\alpha(x,y)=\mu_\alpha f_\beta(x,y)/\mu_\beta$.
Here the last term of $R(\xi,y;\tau)$  that contains the 
function $h_H(x,y)$ is associated with the fluxon
interaction with the borders of the impurity because $h_H(x,y)\neq 0$
only there, i.e., if $(x,y) \notin \partial \Omega$. The first term
corresponds to the radiation produced when the fluxon
passes the bulk of the impurity.

\begin{figure}[htb]
\includegraphics[scale=0.4]{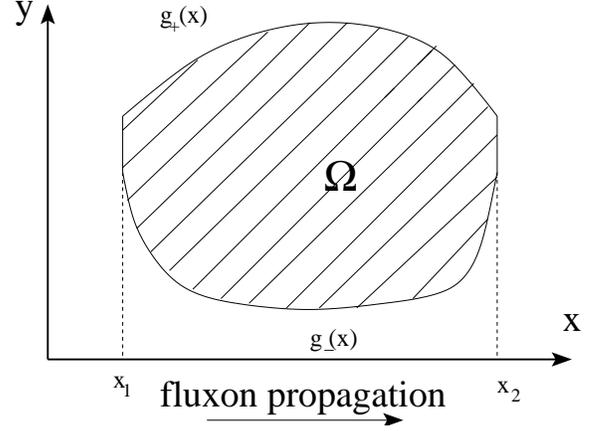}
\caption{Schematic top view of the impurity area $\Omega$.}
\label{fig0}
\end{figure}

The solution of Eq. (\ref{3}) can be represented as
\begin{equation}
\psi(\xi,y,\tau)=\int_{-\infty}^{+\infty}
\int_{-\infty}^{+\infty} a(q_{\xi},q_y;\tau) \varphi(\xi,y;q_\xi,q_y) ~dq_{\xi}~dq_y~,
\end{equation}
where $\varphi(\xi,y;q_\xi,q_y)$ is the eigenfunction \cite{r70jmp,sjs84prb}
of the homogeneous part of this equation:
\begin{eqnarray}
&&\varphi(\xi,y;q_{\xi},q_y)=\frac{e^{i (q_\xi \xi+q_y y)}}{(2\pi)^{3/2}}
\frac{q_\xi+ i \tanh{\xi}}{({1+ q_\xi^2)^{1/2}}},\\
\nonumber
&&\int_{-\infty}^{+\infty} \int_{-\infty}^{+\infty}\varphi^*(\xi,y;q_{\xi},q_y)
\varphi(\xi,y;q_{\xi}',q_y') ~d\xi dy=\\
\label{12}
&=&\frac{1}{2\pi}\delta(q_\xi-q_\xi')\delta(q_y-q_y').
\end{eqnarray}
Here $\delta$ is the Dirac delta function, $q_\xi$ and $q_y$ are the 
components of the plasmon wave vector in the moving frame, and
\begin{equation}\label{12a}
 \bar \omega =\sqrt{1+q_\xi^2+q_y^2}~,
\end{equation}
is the plasmon dispersion law in that frame. The function 
$a(q_{\xi},q_y)$ is the radiation amplitude. It is convenient to 
introduce another function which also describes the emitted radiation, 
namely 
$b(q_\xi,q_y;\tau)\doteq(a_\tau-i\bar \omega a) \exp( i\bar\omega\tau)$.
As a result, the following equality holds:
\begin{equation}
\partial_\tau b=e^{i\bar \omega \tau} \left ( \partial_\tau^2 a+
{\bar \omega}^2a\right )~.
\end{equation}
Multiplying both sides of Eq. (\ref{3}) by 
$\varphi^*(\xi,y;{q}_\xi',{q}_y')$  and integrating simultaneously 
over $y \in\R$ and $\xi \in \R$ we obtain
 $\delta(q_\xi-q'_{\xi})$ and $\delta(q_y-{q}_y')$ on the left-hand side 
 [the orthogonality condition  (\ref{12}) is used] of Eq. (\ref{3}). 
After removing the integration over $q_\xi$ and $q_y$ one arrives at 
the following expression
\begin{eqnarray}\nonumber
\partial_\tau b=2\pi ~e^{i\bar \omega \tau}&&\int_{-\infty}^{+\infty} 
\int_{-\infty}^{+\infty} R(\xi,y;\tau)
 \times\\
&&\times ~ \varphi^*(\xi,y;q_\xi,q_y) ~d\xi~ dy~. \label{13}
\end{eqnarray}
The total radiation over the whole time is defined by the function
\begin{equation}\label{14}
B(q_\xi,q_y)=\int_{-\infty}^{+\infty} \partial_\tau 
b(q_\xi,q_y;\tau) ~d\tau~.
\end{equation}
Thus, with the pair of equations (\ref{13}) and (\ref{14}) one has the 
complete formula for the energy calculation. From this point it is
possible to proceed to the emitted radiation studies for the particular
shapes of $\Omega$. The return to the laboratory frame is performed 
with the help of the following Lorentz transformation:
\begin{eqnarray}\label{16}
q_x=\frac{q_\xi+v\bar \omega}{\sqrt{1-v^2}}, ~
\omega=\frac{vq_\xi+{\bar \omega}}{\sqrt{1-v^2}}~,\\
q_{\xi}=\frac{q_x-v\omega}{\sqrt{1-v^2}}, ~ \label{17}
{\bar \omega}=\frac{\omega-vq_x}{\sqrt{1-v^2}}~.
\end{eqnarray}
The $q_y$ component remains unchanged. Taking into account that the 
emitted energy density equals \cite{m91pd} 
${\cal E}(q_x,q_y)\simeq |B(q_x,q_y)|^2/(4 \pi)$, the total energy
is given by the integral
\begin{equation}\label{15}
E=\int_{-\infty}^{+\infty} \int_{-\infty}^{+\infty} {\cal E}(q_x,q_y)
~d q_x dq_y~.
\end{equation}
The following simplification can be achieved if $\Omega$ has the 
properties
defined below. Suppose the impurity covers the area that is limited
by the lines $x=x_1$ and $x=x_2$ along the $y$ axis and by the
continuous and single-valued functions $y=g_\pm (x)$ along the 
$x$ axis, as is shown in Fig. \ref{fig0}. In this case
\begin{eqnarray}\nonumber
f_{I,H,C}(x,y)&=&\mu_{I,H,C} \left [\theta(x-x_1)-\theta(x-x_2)
\right ]\times\\
&\times& \left \{ \theta [y-g_-(x)]-\theta [y-g_+(x)] \right \}\, ,
\end{eqnarray}
and the integral over $y$ can always be taken. As a result, the 
computation of the radiation function $b$ is reduced considerably. 
Here $\theta(x)$ is the Heaviside function.

Below we consider the concrete examples when the impurity area
$\Omega$ is limited by the piecewise functions.

\subsection{Rectangular impurity}\label{rectum}

In this subsection the rectangular impurity of finite size in both 
$x$ and $y$ directions,
\begin{eqnarray}\nonumber
&&f_{I,H,C}(x,y)={\mu}_{I,H,C} \left [\theta\left (x+\frac{d_x}{2}\right)-
\theta\left (x-\frac{d_x}{2} \right) \right ]\times \\
&&\times \left [\theta \left (y+{d_y \over 2} \right)-\theta \left
(y-{d_y \over 2}\right ) \right ],\label{2}\\
\nonumber
&&h_H(x,y)=\mu_H  \left [\delta \left (x+{d_x \over 2}\right )
-\delta \left (x-{d_x \over 2} \right ) \right ] \times \\
&&\times \left [ \theta \left (y+ {d_y \over 2} \right )-
\theta \left (y-{d_y \over 2} \right ) \right],\label{2a}
\end{eqnarray}
is considered. The parameters $d_x$ and $d_y$ are the impurity length
and width, respectively.

\subsubsection{Spectral density of the emitted waves}

At this point we can substitute the actual expressions
 (\ref{2}) and (\ref{2a}) that corresponds to the rectangular impurity into
Eqs. (\ref{13}) and (\ref{14}). 
Then the radiation function (\ref{14}) in the moving frame is obtained after
the consecutive integration over the $y$, $\tau$, and $\xi$ variables:    
\begin{widetext}
\begin{eqnarray}\nonumber
B(q_{\xi},q_y)&=&i\frac{2\sqrt{2\pi}\mu_I}{q_y^2 \sqrt{1+q_\xi^2}(1-v^2)^{1/2}v^3} 
\sin\left ({q_yd_y \over 2}\right)\sin \left ( \frac{\bar \omega 
\sqrt{1-v^2}}{2v}d_x\right )\mbox{sech}\left [{\frac{\pi}{2v}(q_\xi v+{\bar \omega})}\right ]\times \\
& \times & \left \{ \left (1-v^2-{\mu_H \over \mu_I}+v^2{\mu_C\over \mu_I}\right )
{[{\bar \omega}^2-(1+q_\xi^2)v^2]\over 1+\mu_C}+2{\mu_H \over \mu_I} 
(1-v^2){\bar \omega}^2 \right \}~.
\label{19}
\end{eqnarray}
\end{widetext}
The first term in the curly brackets in Eq. (\ref{19}) appears due to 
the first term in $R$ [see Eq. (\ref{3a})] and can be treated as a 
result of the fluxon interaction with the bulk of the impurity. The 
second term in  the curly brackets appears due to the second term 
(associated  with the function $h_H$) in Eq. (\ref{3a}) and can be 
considered as the radiation that appears  due to the fluxon 
interaction with  the border of  the impurity.
After returning to the laboratory frame of reference with the help of 
Eqs. (\ref{16})-(\ref{17}) the final formula for the 
spectral density reads:
\begin{widetext}
\begin{eqnarray}\nonumber
{\cal E}(q_x,q_y)&=&\frac{2\mu_I^2}{v^4}
\left [\frac{\sin\left({q_yd_y}/{2}\right)}{q_y} \right] ^2
\left \{ \frac{\sin \left[ {d_x(\omega-vq_x)/2v}\right]}{\omega-q_x v} \right\}^2
\mbox{sech}^2 
\left ( \frac{\pi \omega}{2v}\sqrt{1-v^2}\right )\times \\ 
\label{24}
& \times &\frac{\left \{{1-v^2-{\mu_H\over\mu_I}+v^2{\mu_C\over\mu_I}
 \over 1+\mu_C}
\left[(\omega-q_x v)^2+q_y^2v^2\right]+2{\mu_H\over\mu_I}(\omega-vq_x)^2
\right \}^2} {(\omega-q_x v)^2+(v^2-1)q_y^2}~,\\
\omega &=& \sqrt{1+q_x^2+q_y^2}~.
\label{24a}
\end{eqnarray}
\end{widetext}
This function is symmetric with respect to the mirror symmetry
$q_y \to - q_y$ and to the transform $q_x \to -q_x$, $v \to -v$.
Therefore, it is sufficient to restrict the plots of $E(v)$
to the interval $0\le v\le 1$.
In order to compute the total emitted energy $E(v)$
[see Eq. (\ref{15})] it is necessary
to use numerical methods because it is not possible to take 
the respective double integral explicitly.

\subsubsection{1D limit}

Before embarking on the investigation of the full 2D problem it is 
instructive to recall the corresponding
one-dimensional (1D) case of the fluxon scattering on the
impurity with the length $d_x$. Formally this limit can be achieved
if $d_y \to \infty$.  The energy density 
in this case is already known from the previous work \cite{kkc88pla}:
\begin{eqnarray}\nonumber
{\cal E}(q)&=&\frac{\pi}{v^4}\mu_I^2\left (
{1-v^2-{\mu_H\over\mu_I}+v^2{\mu_C\over\mu_I} 
\over 1+\mu_C}+2{\mu_H\over\mu_I}
\right )^2 \times \\
\nonumber
&\times& \sin^2 \left [ \frac{d_x}{2v} 
\left (\sqrt{1+q^2}-q v \right )\right ]\times \\
\label{23}
&\times&  \text{sech}^2\left (\frac{\pi
 \sqrt{1-v^2}}{2 v}\sqrt{1+q^2}\right )~.
\end{eqnarray}
However, in the paper, cited above, the spatial inhomogeneity of
the capacitance was not taken into account. We note that Eq. (\ref{23}) 
can be obtained in the limit $q_y \to 0$ from
Eq. (\ref{24}) ($\mu_I$ should be renormalized as  
$\mu_I d_y \to \mu_I$). This means that the impurity width $d_y$ tends
to infinity, and, as a result, the scattering does not create any
radiation in the $y$ direction, leaving the problem completely
invariant in that direction, i.e., one-dimensional.
   
\begin{figure}[htb]
\includegraphics[scale=0.325]{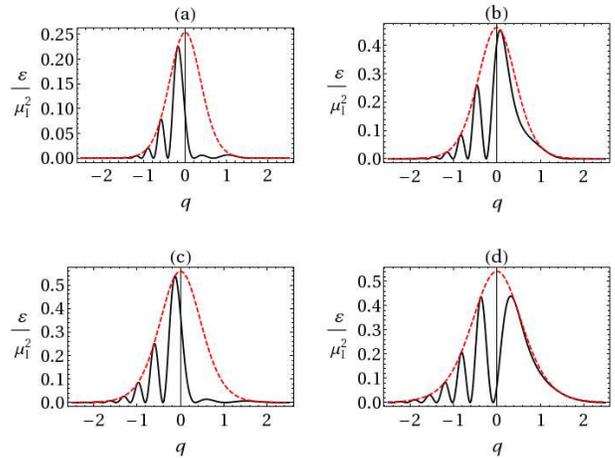}
\caption{(Color online). Energy density [see Eq. (\ref{23})] 
for the 1D junction with $d_x=8$, $\mu_H=\mu_C=0$ at the fluxon
velocity $v=0.398$ (a),
$v=0.488$ (b), $v=0.552$ (c) and $v=0.676$ (d). The red
dashed line depicts the $\mbox{sech}^2$ ``envelope'' term
in Eq. (\ref{23}).} 
\label{figf}
\end{figure}

Typical dependencies of the spectral density ${\cal E}={\cal E}(q)$ 
for the different values of the fluxon velocity are given in 
Fig. \ref{figf}. It is easy to see that the energy density 
${\cal E}(q)$ [Eq. (\ref{23})]  has  an infinite countable set of 
global minima  for which  ${\cal E}(q_{min})=0$. They are  the roots 
of the equation
\begin{eqnarray}\nonumber
&&\frac{d_x (\sqrt{1+q_{min}^2}-q_{min} v)}{2v}=\pi n,~ n=n_0,n_0+1,\ldots~,
\\
&&n_0=\lceil d_x (1-v^2)^{1/2}/(2v\pi) \rceil>0~, \label{25}
\end{eqnarray}
where $\lceil x\rceil$ is the ceiling function \cite{gkp94} of $x$. 
Similarly, there are maxima that are placed between those minima at 
the values of $q$ that are the roots of the equations
\begin{eqnarray}\label{26}
&&\frac{d_x (\sqrt{1+q_{max}^2}-q_{max} v)}{2v}\approx 
{\pi (2n-1) \over 2},~ \\ 
&&n=n_0,n_0+1,\ldots~. \nonumber
\end{eqnarray}
The minima and maxima are associated with the constructive and destructive
interference of the plasmons, emitted when the fluxon enters and
exits the impurity.  Depending on the
length of the impurity and the fluxon velocity, the radiated
plasmons can either cancel each other if their phases differ
by $\pm \pi$  or can enhance each other if their phases coincide.
The radiation consist of the forward ($q>0$) 
and backward ($q<0$) emitted plasmons, and the energy of these
plasmons is distributed non-homogeneously with respect to $q$. 
First of all, the 
most of the energy is concentrated in the long-wavelength modes due to the
presence of the $\mbox{sech}^2(\cdots)$ term in Eq. (\ref{23}). Secondly,
as can be seen from Fig. \ref{figf}, the distribution of the backward 
radiation is defined by the extrema (\ref{25}) and (\ref{26}) that lie
on the negative half-axis ($q<0$). These extrema are distributed almost
in an equidistant way with the step $2\pi v/[d_x(1+v)]$; therefore, the small
change of $v$ will lead to the small change in the area under the ${\cal E}(q)$
curve. On the contrary, the forward radiation 
depends strongly on $v$, especially if $v$ is not small ($v<1$ but not 
$v\ll 1$). Only for large $q$'s the extrema are distributed 
with the almost fixed step  $2\pi v/[d_x(1-v)]$.
The minima of ${\cal E}(q)$ given by Eq. (\ref{25}) come in pairs, 
numbered by the index $n$. These pairs are placed on the different
sides from the value $q=v/\sqrt{1-v^2}$, which is the minimum of the
left-hand side of Eqs. (\ref{25}) and (\ref{26}). The pair with $n=n_0$ is the pair
of the minima, that are the closest to each other. There always should be 
 a maximum between these minima. If the above-mentioned  
minima are very close to each other ($2\pi n_0 v/d_x \gtrsim \sqrt{1-v^2}$), 
the maximum between them cannot be associated with 
Eq. (\ref{26}), as seen in Figs. \ref{figf}(a) and \ref{figf}(c); thus, the
respective value of ${\cal E}$
 lies not on the $\mbox{sech}^2(\cdots)$ envelope function, but
significantly below it. 
%
%
As a result, for these values of
$v$ the forward radiation can be insignificant, as can be observed
from the area below the curve ${\cal E}(q)$ at $q>0$. 
In another case, the pair of minima that correspond to $n_0$ are
significantly separated, and the maximum between them  
belongs to the set (\ref{26}). It is again the first maximum at
the positive axis, and it attains the
value of ${\cal E}$  which is quite large comparing to the
previous case, as can be seen in 
Figs. \ref{figf}(b) and \ref{figf}(d). 

\begin{figure}[htb]
\includegraphics[scale=0.37]{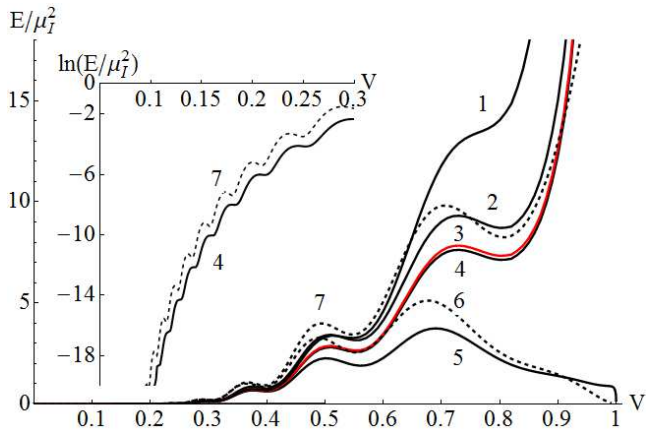}
\caption{(Color online). Total emitted energy (normalized to $\mu_I^2$) 
 as a function of the soliton velocity
for the impurity with $d_x=8$, $d_y=8$ and 
$\mu_H=0$, $\mu_C=0.1$, $\mu_C/\mu_I=1$ (curve 1),
$\mu_H=0$, $\mu_C=-0.05$, $\mu_C/\mu_I=0.5$ (curve 2), 
$\mu_H/\mu_I=0.01$, $\mu_C=0.05$, $\mu_C/\mu_I=0.5$ (curve 3, red),
$\mu_H=0$, $\mu_C=0.05$, $\mu_C/\mu_I=0.5$ (curve 4), and 
$\mu_H=\mu_C=0$ (curve 5). 
The dashed lines 6 and 7 
correspond to the same dependence but for the 1D problem 
[see Eq. (\ref{23}) for the spectral energy density]
 with $d_x=8$ and $\mu_H=\mu_C=0$ (curve 6) 
 and $\mu_H=0$, $\mu_C/\mu_I = 0.5$, $\mu_C = 0.05$ (curve 7). 
 These dependencies are multiplied by a factor $10$ for the sake of 
 convenience. The inset shows the details of the curves 4 and 7.}
\label{fig2}
\end{figure}

The dashed lines $6$ and $7$ in Fig. \ref{fig2} show the dependence of the
total emitted energy on the fluxon velocity (the solid lines correspond
to the 2D case which will be discussed later). 
The values of $v$
which correspond to the minima of the $E(v)$ in 
 line 6 in Fig. \ref{fig2}, have the minimal forward emission, and the 
respective spectral energy distributions are shown in Fig.  \ref{figf}(a), \ref{figf}(c). 
The values of $v$ that 
are the maxima of $E(v)$ correspond to the maximal forward emission
and the respective spectral distributions are given in
Figs. \ref{figf}(b) and \ref{figf}(d). Thus, the maxima of the
total energy coincide with the maximal forward emission while
the minima of $E(v)$ correspond to the minimal forward emission.
It should be noted that the minima [Eq. (\ref{25})] and maxima
[Eq. (\ref{26})] of the energy density are distributed approximately 
equidistantly for the short-wavelength ($|q|\gg 1$) modes but with the
different step for $q>0$ and $q<0$. In the limit  
$|v|\ll 1$ this step becomes approximately the same, it equals
$2\pi v/d_x$. Consequently, in the limit $|v|\to 0$ one cannot
expect sharply pronounced extrema of the $E(v)$ dependence, and this can
be noticed from the inset.

Finally, we remark that in the relativistic
limit $v\to 1$ the total energy $E(v)\to 0$ if $\mu_H=\mu_C=0$ and
$E(v)\to \infty$ if $\mu_{H,C} \neq 0$. The details of this limit will be
discussed below together with the 2D case.

\subsubsection{Total emitted energy in the 2D case}

First of all, we discuss the dependence of the total emitted energy 
$E(v)$ on the impurity parameters $\mu_I$, $\mu_H$ and $\mu_C$. We 
remind the reader that $\mu_I$ is associated with the change of the critical 
current, while $\mu_C$ [see Eq. (\ref{5a})] and $\mu_H$ [see Eq. 
(\ref{6})] appear due to  the narrowing or distension of the 
insulating area. If $\mu_H=\mu_C=0$ the impurity corresponds only to 
the local change of the critical current without any changes in the 
insulating layer thickness. The total emitted energy for the different 
values of  $\mu_H$ and $\mu_C$ is given in Fig. \ref{fig2}. We note the 
principal
difference in the behavior of the $E(v)$ function in the limit 
$v\to 1$ if $\mu_{C,H} \neq 0$ compared to the case $\mu_C=\mu_H=0$.
In the latter case $E(v)$ tends to zero while in the former case
it diverges: $E(v)_{v\to 1}\to +\infty$. The same is observed in the 1D
case (shown by the dashed lines). It is quite obvious from  the
 lines 1 and 4 that for the larger values
of $\mu_C$ the value of the emitted energy is larger. If one takes two
opposite values of $\mu_C$, the case of a microresistor ($\mu_C<0$, line 2) 
yields slightly larger energy emission compared to the case of a
 microshort ($\mu_C>0$, line 4) due to the presence of the $(1+\mu_C)^{-1}$
 coefficient in the energy density (\ref{24}). The effect of the 
 spatial variation of the magnetic field, governed by the coefficient
$\mu_H$ is negligible, as one can observe from the comparison of the
lines 3 and 4. Therefore, we will assume $\mu_H=0$ further on 
throughout the paper.

The divergence  at $v\to 1$ appears due to the presence of the
divergent terms on the right-hand side of Eq. (\ref{3}). These terms 
[see Eq. (\ref{3a})] are proportional to
$(1-v^2)^{-1}$ and $(1-v^2)^{-1/2}$. In the former term the
function ${\bar f}_I$ contains both the parameters
$\mu_C$ and $\mu_H$ and is always finite, thus the divergence
appears only due to the divisor. 
In the latter term, in addition, there is a function $h_H$ which
is non-zero only on the edges of the inhomogeneity, where it is
proportional to the Dirac's  $\delta-$function. 
This term generates the sharp growth of radiation when the fluxon
interacts with the edges of the impurity. In the 1D case it
produces such growth only at the 
entrance ($x=-d_x/2$) and exit ($x=d_x/2$) points of the impurity.
%

We would like to mention that the divergence at $v\to 1$ seems to be non physical. First of all, the presence of the divergent term in 
Eq. (\ref{3a}) means that the first order of perturbation theory is not
applicable any longer in this limit and should be amended somehow. 
Secondly, within the current model the dissipative effects have been 
neglected. If they are taken
into account, the radiated energy will always be finite. 

Other features of the $E(v)$
dependence such as the multiple extrema will be discussed below.
At this point we only note that as $\mu_C$ decreases, the positions of the extrema do not shift significantly, but the absolute values of $E$ at
the extrema decrease. This happens because
the contribution to the emitted radiation due to the 
narrowing/expansion of the insulating layer, decreases. 
Depending on the value of $\mu_C$ some extrema can disappear due to the
growth of $E(v)$ as $v \to 1$ (see line 1 in Fig. \ref{fig2}). The limit
$v\to 0$ is given in the inset of Fig. \ref{fig2}. One can notice
that the extrema of the total energy persist in this limit both in the 1D and
2D cases, although they can be spotted only on the logarithmic scale.

\begin{figure}[htb]
\includegraphics[scale=0.37]{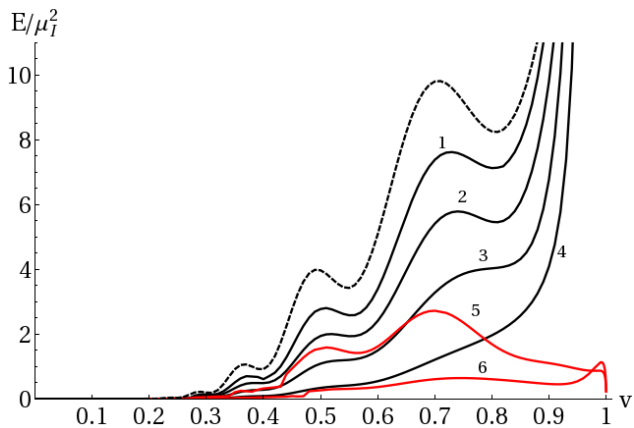}
\caption{(Color online) Total emitted energy (normalized to $\mu_I^2$) 
as a function of the fluxon velocity for 
$\mu_H=0$, $\mu_C/\mu_I=0.5$, $\mu_C=0.05$, $d_x=8$
and $d_y=8$ (curve 1), $d_y=6$ (curve 2), $d_y=4$ (curve 3), and
$d_y=2$ (curve 4). The case $\mu_H=\mu_C=0$ is represented by the red curves
5 ($d_y=6$) and 6 ($d_y=2$).
The dashed line corresponds to the respective 1D problem with
$d_x=8$ (for the sake of convenience it is
multiplied by a factor of 10).
}
\label{fig4}
\end{figure}

\begin{figure}[htb]
\includegraphics[scale=0.32]{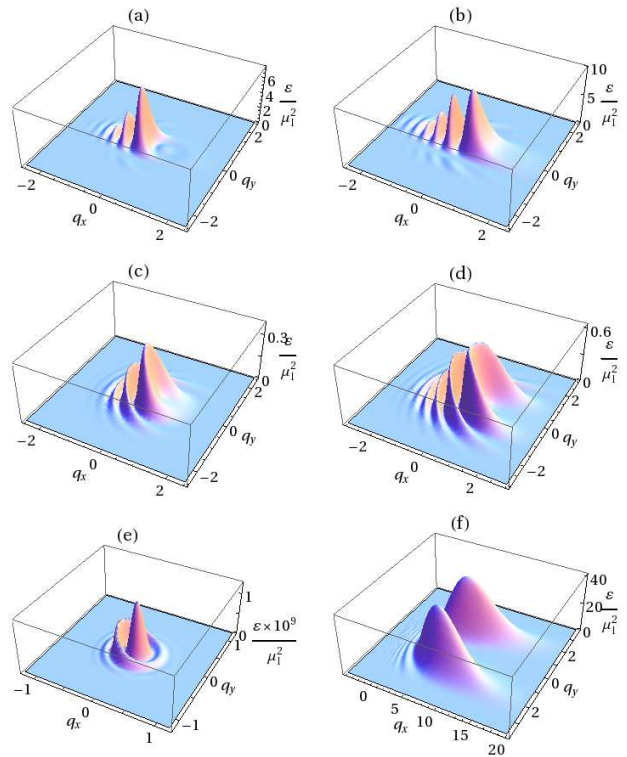}
\caption{(Color online). Emitted energy density ${\cal E}$ for the data in Fig.
\ref{fig4}, curve $1$ at $v=0.55$ (a) and $v=0.73$ (b);
curve $6$ at $v=0.5$ (c), $v=0.75$ (d), $v=0.1$ (e) and $v=0.99$ (f).
}\label{fig5}
\end{figure}

In Fig. \ref{fig4} the
total emitted energy is plotted for the fixed value of the impurity
length $d_x=8$ while its width $d_y$ is varied. The 1D result for the
same length is plotted with the dashed line as a reference. Naturally, 
the value of the emitted energy decreases as $d_y$ decreases. More
interestingly, the extrema become less pronounced, and, finally 
no extrema are seen in curve 4 that corresponds to the case
 $d_y=2$. In the case $\mu_H=\mu_C=0$ we obtain the same picture:
compare curve 5 of Fig. \ref{fig2} ($d_y=8$), curve 5 of 
Fig. \ref{fig4}
($d_y=6$), and curve 6 of Fig. \ref{fig4} ($d_y=2$). The maxima
become more shallow and gradually disappear.  The following 
interpretation of the obtained results can be made. The shape of the 
energy density distribution is given in Fig. \ref{fig5}. The absolute 
minima of the energy density satisfy ${\cal E}(q_x,q_y)=0$ and these 
minimal values are attained at the following set on the $(q_x,q_y)$ 
plane:
\begin{eqnarray}\label{29a}
&& q_y=\frac{2\pi n}{d_y}~,~~n=\pm 1, \pm 2, 
\ldots ~~\mbox{for any}~~ q_x,\\
&& \label{29b}
(1-v^2)q_x^2+q_y^2=\left ( \frac{2\pi m v}{d_x}\right )^2-1
+\frac{4 \pi m v^2}{d_x}q_x, \\
&&~~m=n_0,n_0+1, \ldots~,  \nonumber
\end{eqnarray} 
where $n_0$ is given by Eq. (\ref{25}). Thus, the minima are located
on the set of parallel lines (\ref{29a}) as well as on the set of 
embedded ellipses given by Eq. (\ref{29b}).
The ridges of the maximal ${\cal E}$ lie between the curves, defined by
the roots of Eq. (\ref{29a}). For large $d_y$ these ridges are
strongly localized in the $q_y$ direction [see Figs. \ref{fig5}(a) and \ref{fig5}(b)], 
while the decreasing of $d_y$
makes them concentric and crescent-like as shown in Figs. \ref{fig5}(c) and \ref{fig5}(d).

For the large values of $d_y$ the problem
can be treated as an almost 1D, so that most of the emitted 
radiation travels in the $x$ direction while the $y$- component
of the radiation remains insignificant. 
This can be clearly observed in Figs. \ref{fig5}(a) and \ref{fig5}(b) where the 
spectral density ${\cal E}(q_x,q_y)$ (\ref{24}) is plotted for the values
of velocity close to the minimum [panel a] and maximum [panel b] 
of the curve $1$ in Fig. \ref{fig4}.
Since the decay of the function $[\sin (q_y d_y/2)/q_y]^2$ with the
growth of $q_y$ is quite fast for the large values of $d_y$, the energy density
function remains strongly localized along the $q_x$ axis in the 
neighbourhood of the $q_y=0$ line. Its
behaviour along the $q_x$ axis is reminiscent of the respective 1D
problem, see Eq. (\ref{23}) and Fig. \ref{figf}. Indeed, the minimum
of the total emitted energy corresponds to the minimal forward emission.
It can be easily observed in Fig. \ref{fig5}(a) that the global 
maximum is placed  on the $q_x$ axis at $q_x<0$ 
while the first local maximum at $q_x>0$ is rather small. 
In Fig. \ref{fig5}(b) it can be seen that the global maximum is
placed on the positive half-axis of the $q_x$ axis, and this
happens at $v=0.73$ which is quite close to the maximum of the $E(v)$
function (curve $1$) in Fig. \ref{fig4}.
The further decreasing of $d_y$ smears out maxima in the $E(v)$ 
dependence (compare the curves $1$-$4$ in Fig. \ref{fig5}) up
to the point when only one local maximum can be spotted. The
scattering problem cannot be considered as a quasi-1D any more. 
The radiation distribution becomes rather different as shown 
in Figs. \ref{fig5}(c) and \ref{fig5}(d). The maxima of ${\cal E}(q_x,q_y)$ still
lie on the $q_x$ axis, but the curves (\ref{29b}) that define
the minimal values become distinctly arc shaped. The $y$- component of
the radiation becomes more delocalized and the analogy with the 1D
picture breaks down. 

It is interesting to note how the shape of the energy density function
varies in the extreme limits of the velocity value:
$v\to 0$ and $v\to 1$. 
In the small velocity limit $|v| \ll 1$ the ellipses Eq. (\ref{29b}) that 
correspond to the minima of ${\cal E}$ are almost circles and
the density function is close to being radially symmetric, see
Fig. \ref{fig5}(e). The increasing of $v$ makes the ellipses Eq. (\ref{29b})
more elongated in the $x$ direction, as has been demonstrated previously
[see Figs. \ref{fig5}(a)-(d)].
An interesting situation emerges in the opposite limit, namely if
 $1-|v|\ll 1$. The global maximum that was positioned on the 
 $q_x$ axis splits up into two maxima that are now located 
off the $q_x$ axis symmetrically with respect to each other, as
shown in Fig. \ref{fig5}(f). Physically this means the following. The
slow fluxon ``feels'' the impurity as a wall and the emergent radiation
moves mostly along the fluxon propagation direction. The fast (relativistic)
fluxon interacts with the impurity in such a way that the 
impurity acts like a groin (a wave- breaker) and the emitted
radiation is split by the impurity into two halves that
 have both $x$  and $y$ components. A the same time the $x$ component
 of the radiation becomes insignificant.
 
\begin{figure}[htb]
\includegraphics[scale=0.33]{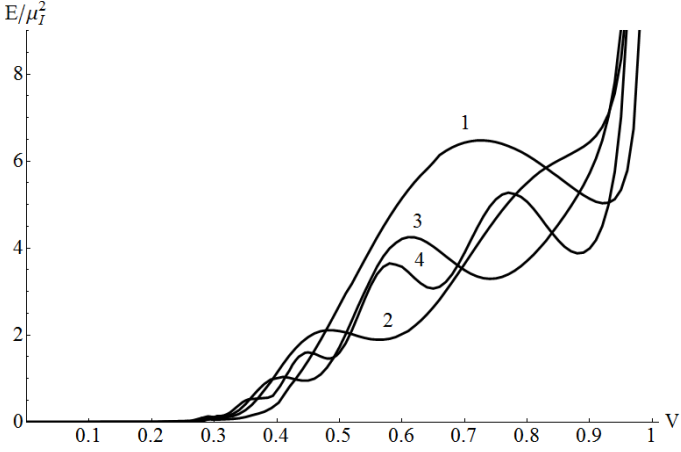}
\caption{Total emitted energy (normalized to $\mu_I^2$) as a function
of the fluxon velocity for $\mu_H=0$, $\mu_C/\mu_I=0.2$, $\mu_C=0.02$, 
$d_y=8$ and $d_x=2$ (curve 1), $d_x=4$ (curve 2), 
$d_x=6$ (curve 3), and $d_x=10$ (curve 4). 
}\label{fig6}
\end{figure}

The number of local extrema of $E(v)$ depends on the impurity 
length $d_x$. This is easily
demonstrated by Fig. \ref{fig6}, where the number of the maxima 
decreases with the decreasing of $d_x$. This result is similar to the
same situation in the 1D model \cite{kkc88pla}.

\subsubsection{Limiting cases}

It is of interest to check the limiting cases when in one of the
directions  ($x$ or $y$) the impurity becomes infinitely narrow.
 In the first case the limit $d_x \to 0$, while $\mu_*=\mu_I d_x$ remains 
 constant, corresponds to the situation when the impurity becomes
 infinitely thin in the $x$ direction. In Eqs. (\ref{2}) and (\ref{2a}) 
 the difference of 
the $\theta$-functions that form the first factor in
the functions $f_{I,H,C}(x,y)$ (\ref{2}) becomes the Dirac $\delta$ function
while $h_H \equiv 0$. This case with $\mu_C=\mu_H=0$ has been studied 
previously \cite{sz13ujp}. 
Yet another interesting limit can be considered if $d_y\to 0$,
$\mu_*=\mu_I d_y$. In other words, the impurity remains elongated in the
$x$ direction, but becomes infinitely thin in the $y$ direction.
The spectral density of the emitted plasmons in the cases mentioned
above reads
\begin{widetext}
\begin{eqnarray}\label{29c}
\frac{{\cal E}(q_x,q_y)}{\mu_*^2}\to \left \{\begin{array}{ccc}
\displaystyle
&&\displaystyle
\frac{\sin^2 \left ( \frac{q_y d_y}{2}\right )}{2v^6q_y^2}
\frac{\left \{{1-v^2-{\mu_H\over\mu_I}+v^2{\mu_C\over\mu_I}\over{1+\mu_C}}
\left[(\omega-q_x v)^2+q_y^2v^2\right ]+2{\mu_H\over\mu_I}(\omega-vq_x)^2
\right \}^2}{(\omega-q_x v)^2+(v^2-1)q_y^2}\times\\
&&~\times \text{sech}^2\left(\frac{\pi\omega \sqrt{1-v^2}}{2 v}\right)~,~~\textrm{if}~
d_x \to 0,\\
\label{29d} &&
\displaystyle 
\frac{1}{2v^4} \left [\frac{\sin [d_x(\omega-q_xv)/2v]}{\omega-q_xv} \right ]^2
\displaystyle
\frac{\left \{{1-v^2-{\mu_H\over\mu_I}+v^2{\mu_C\over\mu_I} \over 
{1+\mu_C}}\left [(\omega-q_x v)^2+q_y^2v^2\right ]+2{\mu_H\over\mu_I}(\omega-vq_x)^2\right \}^2}{(\omega-q_x v)^2+(v^2-1)q_y^2}\times\\
&&~\times  \text{sech}^2\left(\frac{\pi\omega \sqrt{1-v^2}}{2 v}\right)\,,
~~~ \textrm{if}~ d_y \to 0~.
\end{array} \right .
\end{eqnarray}
\end{widetext}
In these limits the modulation in $q$- space, caused by the 
interference,  disappears along the $q_x$ direction in the first 
formula because the impurity length becomes infinitely small. For the 
same reason there is no interference along the $q_y$ component when 
$d_y\to 0$ in the second formula of Eq. (\ref{29d}). 
When any of these limits are approached, the multiple maxima of the 
$E(v)$ dependence disappear leaving only one local maximum. 
The limit of the point [$f_I(x,y)=\mu_I\delta(x)\delta(y)$] impurity \cite{m91pd} 
can be achieved easily
from the stripe impurity by taking in Eqs. (\ref{29c}) the limits 
$\mu_{C,H}\to 0$ and $d_x\to 0$ or $d_y \to 0$ where appropriate. When the impurity shrinks into a point the local change of 
the insulating layer thickness is ignored; thus $\mu_C=\mu_H=0$.
The obtained formula coincides with the previous result \cite{m91pd}.

\subsection{Rhombic impurity}\label{rhomb}

Now we consider the rhombus(diamond)-shaped impurity with  
$d_x$ and $d_y$ being its length and width respectively:
\begin{equation}
\Omega:~~|x|\le d_x/2~ \bigcap ~|y|\le g(x)  =d_y(1/2-|x|/d_x)~.
\end{equation}
The tip of the rhombus is perpendicular to the fluxon line. Then
\begin{eqnarray}\nonumber
h_H(x,y)&=&\mu_H \left \{ \left [\delta(x+d_x/2)-\delta(x-d_x/2) \right ]\times \right . \\
\nonumber
&\times & \left .\left [ \theta(y+g(x))-\theta(y-g(x)) \right]\right .+ \\
\nonumber
&+& \left .
\left [ \theta (x+d_x/2)-\theta(x-d_x/2)\right ] \right . \times \\
\label{29}
&\times& \left .\left \{ \delta [y-g(x)]+\delta [y+g(x)] \right \} g'(x)\right \},\\
g'(x)&=&-\frac{d_y}{d_x}\mbox{sign}(x)=-\frac{d_y}{d_x}
\mbox{sign}(\xi+v\tau)~.\label{30}
\end{eqnarray}
Substituting the formulas (\ref{29}) and (\ref{30}) into Eqs. (\ref{13}) and 
(\ref{14}) we obtain the radiation function in the moving frame:
\begin{widetext}
\begin{eqnarray}\nonumber
B(q_{\xi},q_y)&=&i\frac{2\sqrt{2\pi}\mu_I}{q_y^2 \sqrt{1+q_\xi^2}(1-v^2)^{1/2}v^3}{d_x \over d_y}
\left \{ \left (1-v^2-{\mu_H \over \mu_I}+v^2{\mu_C\over \mu_I}\right ){[{\bar \omega}^2-(1+q_\xi^2)v^2]\over {1+\mu_C}}+2{\mu_H \over \mu_I} (1-v^2){\bar \omega}^2 \right \}\times \\
& \times & \frac{\cos \left (\frac{q_yd_y}{2} \right )-
\cos \left( \frac{\bar \omega \sqrt{1-v^2}}{2v}d_x\right)}
{({d_x \over d_y}{\bar \omega})^2\frac{1-v^2}{q_y^2 v^2}-1}
\mbox{sech}\left [{\frac{\pi}{2v}(q_\xi v+{\bar \omega})}\right ]~,
\label{31xx}
\end{eqnarray}
\end{widetext}
where the dispersion law $\bar \omega=\bar \omega(q_\xi,q_y)$ in the 
moving frame is given by Eq. (\ref{12a}). 
The transition to the laboratory frame is performed in the 
standard way, and- as a result, the spectral energy density in the 
laboratory frame is expressed by the following formula:
\begin{widetext}
\begin{eqnarray}\nonumber
{\cal E}(q_x,q_y)&=&\frac{2\mu_I^2}{v^2}\left (\frac{d_x}{d_y} \right)^2 
\left \{\frac{\cos(q_yd_y/2)-\cos [d_x(\omega-q_xv)/2v]}
{\left (\frac{d_x}{d_y} \right)^2(\omega-q_x v)^2-v^2 q_y^2} \mbox{sech} 
\left ( \frac{\pi \omega}{2v}\sqrt{1-v^2}\right )\right \}^2
\times \\
&\times & 
\frac{\left [{1-v^2-{\mu_H\over\mu_I}+v^2{\mu_C\over\mu_I}
\over(1+\mu_C)}\left((\omega-q_x v)^2+q_y^2v^2\right)+2{\mu_H\over\mu_I}
(\omega-vq_x)^2\right ]^2}{(\omega-q_x v)^2+(v^2-1)q_y^2}~,
\label{32}
\end{eqnarray}
\end{widetext}
where the dispersion law $\omega=\omega(q_x,q_y)$ in the laboratory 
frame is given by Eq. (\ref{24a}). It may seem that this dependence
has a singularity where the equation 
$
\left (\frac{d_x}{d_y} \right)^2(\omega-q_x v)^2=v^2 q_y^2
$
is satisfied. However, with the help of the trigonometric
formula $\cos a - \cos b=2 \sin [(a+b)/2]\sin [(b-a)/2]$ it is
straightforward to show that the respective divergences cancel out.  

\begin{figure}[htb]
\includegraphics[scale=0.33]{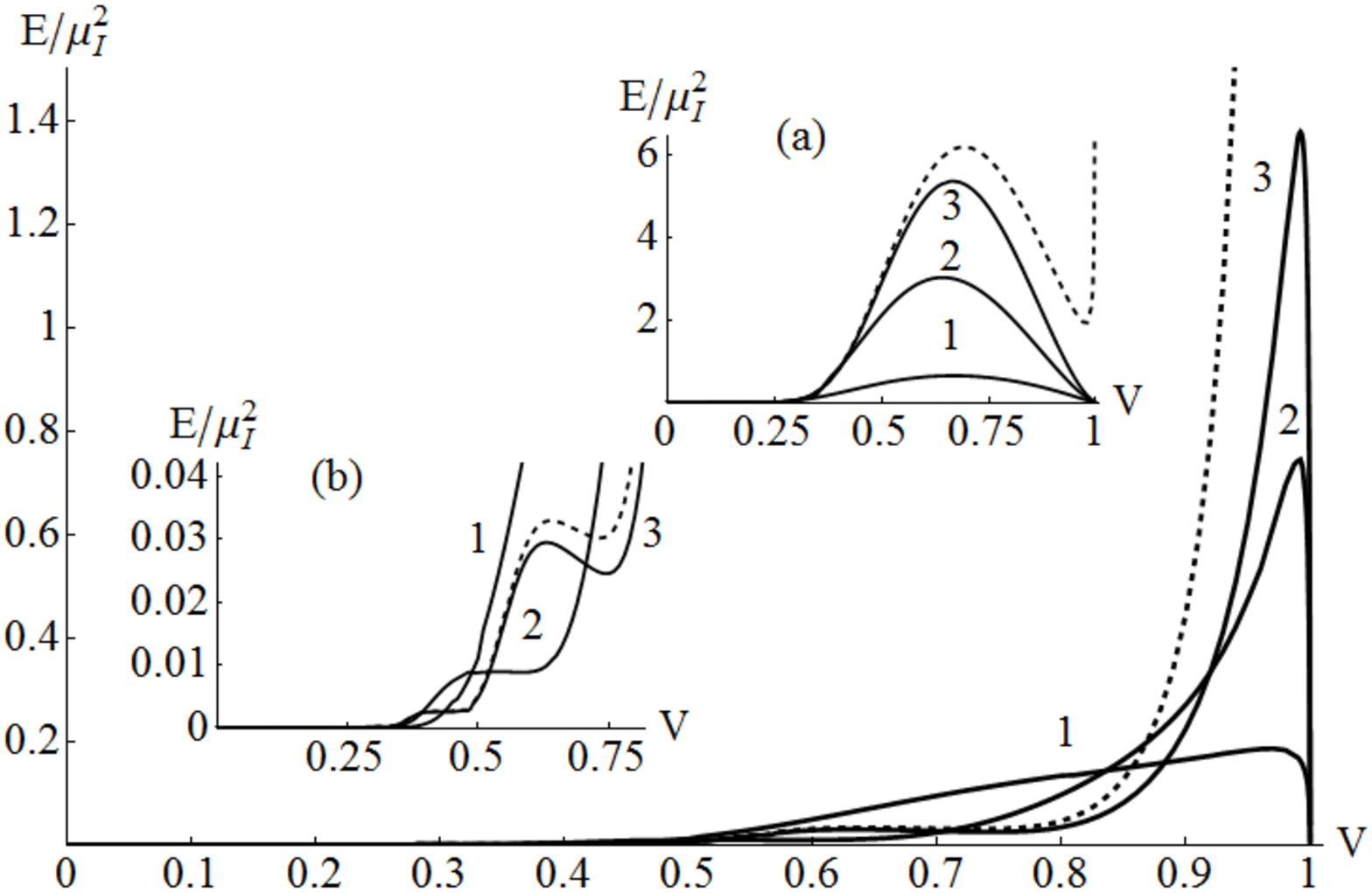}
\caption{Total emitted energy (normalized to $\mu_I^2$) as a function
of the fluxon velocity for the rhombic impurity 
 with the fixed ratio $d_x/d_y=4$. The solid lines correspond to
 $\mu_C=\mu_H=0$, 
$d_x=4$ (curve 1), $d_x=8$ (curve 2) and
$d_x=12$ (curve 3). The dashed line corresponds to $d_x=12$ and $\mu_C=0.01$
and $\mu_C/\mu_I=0.1$, $\mu_C=0.01$, $\mu_H=0$. 
The inset (a) corresponds to the case $d_x/d_y=1/4$, $\mu_C=\mu_H=0$ $d_x=1$ (curve 1), 
$d_x=2$ (curve 2) and $d_x=3$ (curve 3) and $\mu_C/\mu_I=0.1$, 
$\mu_H=0$, $d_x=3$ (dashed curve).
The inset (b) shows the details of the main figure.
}
\label{fig7}
\end{figure}

\begin{figure}[htb]
\includegraphics[scale=0.4]{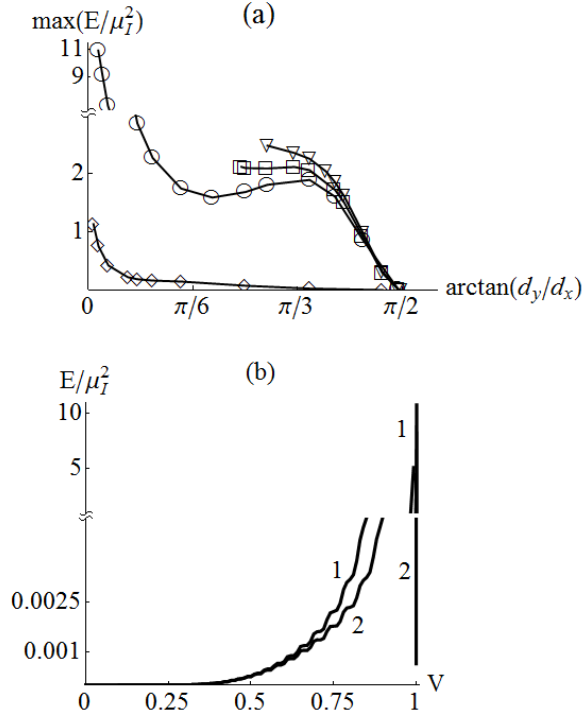}
\caption{
(a) The value of the local maximum $\max_{v \in [0,1[}E(v)$
of the emitted energy (normalized to $\mu_I^2$) as a function
of the angle $\arctan(d_y/d_x)$ 
 for the parameters $d_y=5$, $\mu_C=0.01$, $\mu_C/\mu_I=0.1$ ($\nabla$), 
 $\mu_C=0.005$, $\mu_C/\mu_I=0.05$ ($\Box$), $\mu_C=0$ ($\circ$) and
 $d_y=1$, $\mu_C=0$ ($\diamond$). 
 The solid line is used as a guide for an eye. \\
 (b) Emitted energy dependence  (normalized to $\mu_I^2$)
as a function of the fluxon velocity for $d_x=100$, $d_y=5$, 
$\mu_C=0.01$, $\mu_C/\mu_I=0.1$ (curve 1) and 
$\mu_C=0$ (curve 2). $\mu_H=0$ everywhere.
}\label{fig8}
\end{figure}

\begin{figure}[htb]
\vspace{0.5cm}
\includegraphics[scale=0.3]{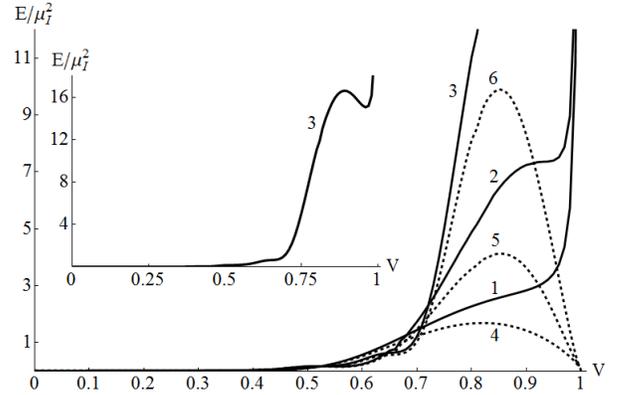}
\caption{Total emitted energy (normalized to $\mu_I^2$) as a function
of the fluxon velocity for the square rhombic ($d_x=d_y$) impurity 
at $\mu_H=0$, $\mu_C=0.01$, $\mu_C/\mu_I=0.1$, $d_x=d_y=5$ (curve 1), 
$d_x=d_y=10$ (curve 2),  $d_x=d_y=20$ (curve 3).
 The dashed lines corresponds to the case $\mu_H=\mu_C=0$ and 
 $d_x=d_y=5$ (curve 4), $d_x=d_y=10$ (curve 5),
 $d_x=d_y=20$ (curve 6).
The inset gives the details of curve 3 on the larger scale.}
\label{fig9}
\end{figure}

The total emitted energy as a function of the fluxon velocity $v$ 
is shown in Figs. \ref{fig7}-\ref{fig9}. The first figure (Fig. 
\ref{fig7}) focuses on the situation when the ratio $d_y/d_x$ is 
fixed while the area covered by the impurity is varied. The main figure
correspond to the impurity with its narrow edge pointing towards
the fluxon direction ($d_x/d_y=4$). The inset (a) describes the 
opposite situation: $d_y/d_x=4$. In general, 
the dependence $E(v)$  grows with $v$ in the limit $v\ll 1$ and 
diverges at $v\to 1$ due to the presence of the $\mu_C$ and $\mu_H$ 
terms [otherwise, if $\mu_C=\mu_H=0$, we have $E(v)_{v\to 1}\to 0$]. 
This behavior is quite similar to the case of the rectangular 
impurity studied in \ref{rectum}. 

First we consider the rhombus, elongated towards the fluxon propagation 
direction (main part of Fig. \ref{fig7}). We observe that in the case 
$\mu_C=\mu_H=0$ there is one well- established maximum of the $E(v)$ 
dependence  which is positioned very close to the value $v=1$.
As the impurity area is increased, the peak of the energy dependence sharpens,
while the position of the maximum shifts towards the point $v=1$.
If $\mu_C\neq 0$ the main maximum disappears due to the unbounded 
growth of the energy dependence. There are other maxima of the $E(v)$
dependence, however they are very weak and can be noticed only if
the respective region is zoomed [see the inset (b)]. When the impurity
area decreases, some of these maxima disappear [compare the curves 3
and 2 in the the inset (b)].
   

Inset (a) of Fig. \ref{fig7}  corresponds to the situation
when the impurity 
is elongated in the $y$- direction with the ratio $d_x/d_y=1/4$
being fixed. In this case there is only one local maximum that
decreases while the adjacent local minimum  
becomes more shallow as the area $d_xd_y/2$ decreases. This case is 
qualitatively close to the limit of the strip impurity
\cite{sz13ujp} but the limit (\ref{29c}) is not restored mathematically.

It is possible to consider the limiting cases of the 
infinitely narrow stripes: $d_x\to 0$ and $d_y\to 0$.
If the impurity amplitude is redefined as $\mu_*=\mu_I d_x$ (or 
$\mu_*=\mu_I d_y$), the spectral density in these limits reads
\begin{widetext}
\begin{eqnarray}
\frac{{\cal E}(q_x,q_y)}{\mu_*^2}\to \left \{\begin{array}{ccc}
\displaystyle
\label{31}&&\displaystyle
8\frac{\sin^4 \left ( \frac{q_y d_y}{4}\right )}{v^6 d_y^2 q_y^4}
\frac{\left \{{1-v^2-{\mu_H\over\mu_I}+v^2{\mu_C\over\mu_I}
\over 1+\mu_C}\left [(\omega-q_x v)^2+q_y^2v^2\right ]+2{\mu_H\over\mu_I}
(\omega-vq_x)^2\right \}^2}{(\omega-q_x v)^2+(v^2-1)q_y^2}\times\\
&&~\times ~ \text{sech}^2\left(\frac{\pi\omega \sqrt{1-v^2}}{2 v}\right)~, ~~\textrm{if}~d_x \to 0,\\
\label{31x} &&
\displaystyle 
{{8}\over{d_x^2v^2}}\left \{\frac{\sin [d_x(\omega-q_xv)/4v]}{\omega-q_xv} 
\right \}^4
\frac{\left \{{1-v^2-{\mu_H\over\mu_I}+v^2{\mu_C\over\mu_I}
\over 1+\mu_C}\left [(\omega-q_x v)^2+q_y^2v^2\right ]+2{\mu_H\over\mu_I}(\omega-vq_x)^2\right \}^2}{(\omega-q_x v)^2+(v^2-1)q_y^2}\times\\
&&~\times  \text{sech}^2\left(\frac{\pi\omega \sqrt{1-v^2}}{2 v}\right)\, ~~~ \textrm{if}~ d_y \to 0~.
\end{array} \right .
\end{eqnarray}
\end{widetext}
These limiting values of ${\cal E}$ are very similar to the analogous
limits for the rectangular impurity (\ref{29c}). The only principal 
difference is the interference terms that are responsible for the
oscillations in the $q_x$ or $q_y$ direction come with the power $4$
and not $2$ as in Eq. (\ref{29c}). 

Next we
focus on the situation when the impurity width $d_y$ is fixed and
its length $d_x$ is varied. In Fig. \ref{fig8}(a) the dependence
of the local maximum value (defined within the interval $0\le v <1$)
of the emitted energy 
as a function of the rhombus angle $\arctan(d_y/d_x)$ is plotted.
If $\mu_C=\mu_H=0$ the $\max_{v \in [0,1[}E(v)$ dependence on the
rhombus angle is a decaying function almost everywhere in the
interval $[0,\pi/2]$. In the limit $d_x\to \infty$ the maximum of
$E(v)$ grows as the amount of the emitted energy increases. 
Only in the neighborhood of the angle 
$\pi/3$ there is a weakly pronounced local maximum. If $\mu_C \neq 0$
such a dependence cannot be defined for the whole interval $[0,\pi/2]$
and it starts from some critical value of the angle (see the
dependencies, marked by squares and inverted triangles) and continues till the
value $\pi/2$. 
Below this critical
angle there is no local maximum of $E(v)$ because it 
becomes strictly monotonic. 
If $d_y$ is decreased, the dependence becomes a 
strictly decaying function (shown by the circles in Fig. \ref{fig8}) that
cover the whole interval $[0,\pi/2]$ even if $\mu_C\neq 0$.
In Fig. \ref{fig8}(b) the $E(v)$ dependence is demonstrated in the
limit of the extremely narrow rhombic impurity. If $\mu_C>0$ there
is no maximum and the $E(v)$ function is a monotonically increasing function.
If $\mu_C=0$ there is a sharp maximum very close to $v=1$ and everywhere
else the function behaves almost identically to the case $\mu_C>0$.
One can notice a fine structure of multiple inflection points. These
points are the remnants of the local maxima that are clearly seen
in the inset (b) of Fig. \ref{fig7}. The number of these inflection points
increases as the length of the rhombus $d_x$ increases.
Here we observe a weak link with the case of the rectangular impurity,
studied in \ref{rectum}. In that case we reported the
increasing of the number of maxima of $E(v)$ when $d_x$ increased. 
For the rhombus we see the maxima degenerate into the inflection points.
The limit $d_x\to\infty$ means that the impurity acts as
an extremely narrow groin that does not cause much radiation
due to its narrowness for small and intermediate velocities. Significant
growth of the emitted radiation can be spotted only in the relativistic
regime ($1-v^2\ll 1$).
It is important to remark that there is no clear 1D limit for the rhombic
impurity, while such a limit can be achieved for the rectangular impurity
by setting $d_y\to\infty$.

In the limits $d_x \to 0$ the radiated energy decreases significantly as 
one obtains infinitely thin impurity
in the $x$ direction. When this limit is approached the local maximum
of the $E(v)$ dependence becomes less and less pronounced.
The energy density is proportional
to $d_x^2$; thus, it is not surprising that the total energy
tends to zero in this limit.  The renormalization of the
impurity amplitude $\mu_*=\mu_I d_x$ and $d_x\to 0$
will lead  the first formula of Eq. (\ref{31}).

If the rhombus becomes a square ($d_x=d_y$)  the local maximum of the 
radiation becomes more pronounced
if the area of the impurity increases, as shown in Fig. \ref{fig9}. 
Also, the decreasing 
of the impurity area makes the local maximum less pronounced.
The main maximum is dominant, although there exist  
 secondary local maxima, to the left from the main maximum, 
although they are very small. The position of the main maximum 
shifts to the left as the impurity size is decreased; however this
shift is insignificant even if the area $d_xd_y/2$ is decreased
by the order of magnitude (compare the curves 4 and 6 in Fig. \ref{fig9}).
Reducing the size of the impurity in both directions ($d_x,d_y\to 0$, 
and $\mu_{C,H}\to 0$) brings the spectral energy density function 
(\ref{32}) to the already known limit of the point-like 
impurity\cite{m91pd}. The same
limit can be obtained from any of the Eqs. (\ref{31}) by setting
$d_y\to 0$, $\mu_{C,H} \to 0$ in the first equation or 
$d_x\to 0$, $\mu_{C,H} \to 0$ in the second equation. The impurity 
amplitude should be redefined as $\mu_*=\mu_I d_y$ or $\mu_*=\mu_I d_x$, respectively. 

\begin{figure}[htb]
\includegraphics[scale=0.33]{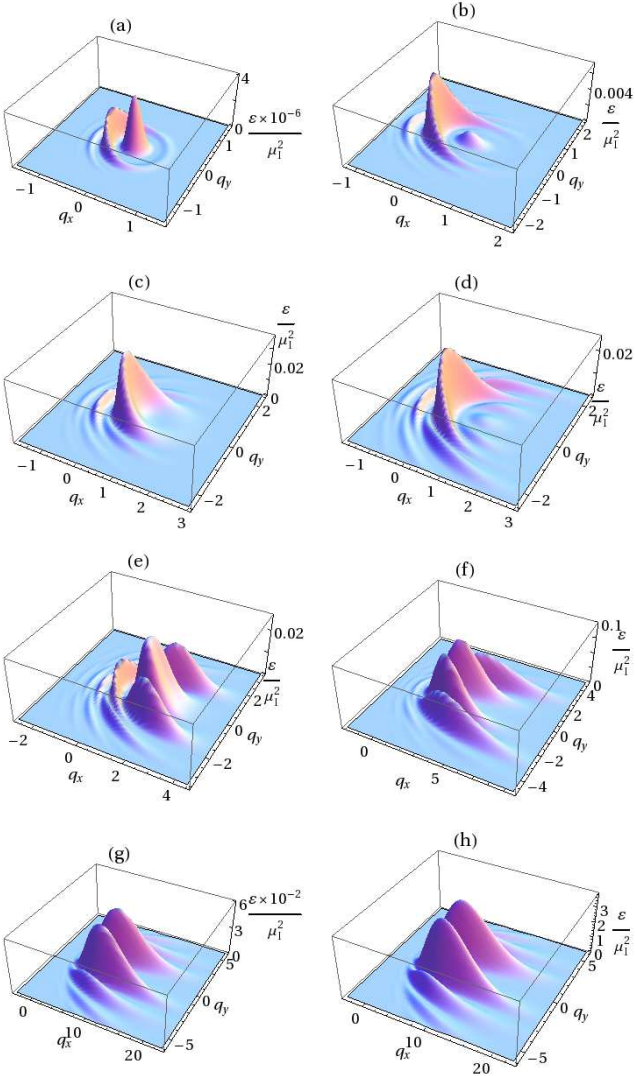}
\caption{(Color online). Emitted energy density ${\cal E}$ for 
the rhombic impurity with $d_x=12$, $d_y=3$, $\mu_C=\mu_H=0$ 
(curve 3 in Fig. \ref{fig7}) at $v=0.2$ (a), $v=0.47$ (b),
 $v=0.63$ (c), $v=0.75$ (d), $v=0.85$ (e), $v=0.95$ (f), and 
 $v=0.993$ (g). The panel (h) 
 corresponds to $v=0.993$ and $\mu_C=0.01$,  $\mu_C/\mu_I=0.1$.
}\label{fig10}
\end{figure}

The energy density profiles ${\cal E}(q_x,q_y)$ that correspond to the 
rhombic impurity are presented in Fig. \ref{fig10}. As a particular 
example, we consider an impurity that corresponds to the curve $3$ from 
Fig. \ref{fig7}, i.e., for $d_x=12$, $d_y=3$. 
 This energy density distribution bears many qualitative
similarities with the energy density function for the
rectangular impurity shown in Fig. \ref{fig5}.  
The global minima of the energy density satisfy the condition
${\cal E}(q_x,q_y)=0$ and are given by the set of equations
\begin{equation}\label{36}
d_x\left ( \frac{\sqrt{1+q_x^2+q_y^2}}{v}-q_x\right)\pm q_y d_y=
4\pi n_{\mp}~.
\end{equation}
This set of equations describes the sequence of pairs of 
ellipses that are numbered by the integers $n_\pm$, 
\begin{equation}
n_\mp=n_0,n_0+1,\ldots,~
n_0=\left\lceil \frac{d_x}{4\pi}\sqrt{\frac{1}{v^2}-
\frac{d_x^2+d_y^2}{d_x^2}} \right\rceil~.
\end{equation}
if 
\begin{equation}\label{40}
|v|< \frac{d_x}{\sqrt{d_x^2+d_y^2}}~.
\end{equation}
Otherwise, the Eqs. (\ref{36}) yield the set of hyperbolas that are
numbered with $n_{\pm}=\pm 1, \pm 2, \ldots$.
The two curves (ellipses or hyperbolas) given by Eq. (\ref{36}) that
correspond to 
the opposite signs but with $n_+=n_-$
are mapped into each other with the mirror symmetry with respect 
to the $q_x$ axis. If we consider the set of curves with the
same sign, say $+$, they are embedded into each other and they
expand with the growth of the index $n_+$.
Between these curves lie the ridges of the ${\cal E}(q_x,q_y)$ function,
and the local maxima of the energy density lie on these ridges.

The signatures of these curves can be spotted in all panels of
Fig. \ref{fig10}. 
For  small and intermediate values of the fluxon velocity 
 the emitted radiation is localized predominantly in one peak in 
 the $q-$space, as shown in Figs. \ref{fig10}(a-d). This peak lies 
 on the $q_x$ axis; thus, most of the radiation does not propagate 
 in the perpendicular direction. In the panel (a) one can observe
the distribution for the rather small value of the fluxon velocity
($v=0.2$) and this distribution is close to being radial.
At such small velocities the pair of ellipses (\ref{36}) 
with $n_-=n_+$ are very close to being circles and 
almost coincide with each other. For larger values of $v$
these pairs start to separate, as illustrated in Figs. \ref{fig10}(b-e).
The panel (b) corresponds to the local minimum 
 of $E(v)$ (curve $3$ of Fig. \ref{fig7}) at $v=0.47$ while the 
 panel (c) corresponds to the local maximum at $v=0.63$. 
The structure of both these
 functions is similar and the only difference is that the
maximal peak in panel (c) lies in the area of backward 
radiation ($q_x \approx -0.25$), while 
in panel (c) the main 
peak lies on the positive half of the $q_x$ axis at $q_x\approx 0.3$.
Thus, for the intermediate velocities the situation is similar
to the case of rectangular impurity, where the minimum of $E(v)$ 
corresponded to the minimal forward radiation. 
Panel (d) corresponds  to the next 
local minimum of the $E(v)$ curve at $v=0.75$, and here one observes
the increasing of the share of the perpendicular radiation in the total
radiated energy. The further increasing of $v$ leads to the 
appearance of the pair of equivalent
local maxima off the $q_y=0$ axis [see panel (e)]. These maxima
become global as $v$ approaches the value $v=1$ [see panels (f) and (g)].
Thus, we observe the increasing of the perpendicular
radiation that reaches its climax in the relativistic limit
$v\to 1$. Panel (f) corresponds to the maximum
of the $E(v)$ function (curve $3$ of Fig. \ref{fig7}) at
$v=0.993$. According to Eq. (\ref{40}) the minima of the energy 
density lie on the hyperbolas and the maxima lie between these
hyperbolas and off the $q_x$ axis. They appear to be strongly
localized in the $q_y$ direction while their localization in the
$q_x$ is significantly weaker.
In this limit the interaction time with the tip of the rhombus  
is too small to generate significant 
longitudinal radiation, and the shape of the obstacle breaks
the incident fluxon as a groin and generates predominantly 
transverse radiation. 

Finally, we mention the dependence of the emitted energy on the
parameter $\mu_C$. Panel (h) corresponds to the same
parameters of the model as in panel (g) but with $\mu_C>0$.
Comparing panels (g) and (h) we see that the structure of these
functions is very similar while the absolute values of ${\cal E}$
are significantly smaller in the $\mu_C=0$ case. 
If $\mu_C=0$,  but for the same value of the
fluxon velocity, the values of the maxima actually decrease with 
$v$. Thus, the total emitted energy tends to zero, in the same way as 
shown by the dashed lines in Fig. \ref{fig9}. This has been confirmed 
for the values of $v$ even closer to unity as well as for the different
values of $d_{x,y}$. The qualitative behavior of $E(v)$ in the
limit $|v|\to 1$ appears to be the same both for the 
rectangular and rhombic impurities.

\section{Discussion and conclusion}\label{discuss}

The radiation emitted as a result of the fluxon interaction with the
impurity of a general geometrical shape in the large two-dimensional
Josephson junction has been studied. The emitted energy distribution
in the $q-$space has been computed as well as the total emitted energy.
This energy distribution can always be represented as a triple integral.
In principle, any geometrical shape can be taken into consideration; 
however, the explicit integration is not always possible, but if the inhomogeneity area can be represented by the piecewise-linear functions, 
this integration can be done. In this article the rectangular and 
rhombic impurities have been studied. 

The main result of this work has been formulated in the 
dependence $E(v)$ of the total emitted energy as a function of the
incident fluxon velocity.  It appears that this dependence
has local maxima that depend strongly on the geometric properties
of the impurity. These local maxima do not exist if the impurity
is treated as a point\cite{m91pd}. Controlling the shape of the
impurity one can remove the extrema or make them more pronounced.
The limit of the 1D problem with the finite-size \cite{kkc88pla} 
inhomogeneity can be restored.

First of all we would like to mention the differences between the 1D
and 2D cases. The 1D case appears to be the limit of the 2D
rectangular impurity case when $d_y \to \infty$. While moving away
from the 1D limit by decreasing $d_y$ we observe gradual lowering and
disappearance of the extrema of the $E(v)$ dependence. Next, the 2D
model allows to take into account the impurity shapes that are different
from the rectangle. For the rhombic impurity we have demonstrated that
the emitted energy dependence on the fluxon velocity is rather
different from the rectangular case and does not possess the 1D limit.
In principle, other geometrical impurity shapes can be studied, 
including the asymmetric ones. 

In this article the junction thickness change due to the
homogeneity is taken into account. Its role is measured by the
parameters $\mu_H$ and $\mu_C$ [see Eqs. (\ref{3b}) and (\ref{6})]. 
The parameter $\mu_C$ is responsible for the capacitance change and
plays the dominant role. In some 
papers \cite{bs95prb} these parameters are ignored (especially they are always
ignored if the point impurities are considered), and, 
in general, are considered
to be weak \cite{kkc88pla}. However, the junction thickness change influences significantly 
the asymptotic behavior of the total emitted energy in the ``relativistic'' 
(i.e., $v\to 1$) limit of the fluxon velocity. If the thickness change
is ignored, the total emitted energy goes to zero, while it exhibits
unbounded growth if the thickness change is taken into account.
This is true for the both 1D and 2D junctions. The emitted energy
has been computed under the assumption that it is a small perturbation
on the fluxon background. Consideration of the higher order corrections
may block the infinite radiation growth. Also, the dissipative effects,
which have been ignored in this work, should contribute to the
 decreasing of the emitted energy. 

Although the real large-area Josephson junctions have finite
dimensions, in this article the infinitely-sized junction has been
considered. This approximation is sufficient if the physical dimensions
of the LJJ exceed by the order of magnitude the Josephson penetration
depth and, consequently, the fluxon length in the $x$ direction.
The boundary conditions are also important, 
however \cite{loes85jap,elos85jap}, if the junction width is large enough
(exceeds the Josephson length at least by the order of magnitude)
the fluxon distortion from the linear shape is insignificant.
In any case, before focusing on the more concrete setup an
idealized, but more easily solvable model should be studied.

Finally, we discuss the possible application of the obtained 
results. Recently, a number of papers have focused on the
different application of the fluxon dynamics in the 2D LJJ, such
as fluxon splitting on the T-shaped junctions \cite{gk06prl},
excitation of the different modes that move along the fluxon
front \cite{gksyn08prl,gksyn09prb}, and the fluxon logic gates \cite{nk10pc}
where the interaction with the spatial inhomogeneity takes place. 
If the incident fluxon velocity is large enough, the emitted
radiation becomes sufficient and it should influence the fluxon motion.
In particular,  the non-monotonicity of the $E(v)$ dependence may
produce the hysteresis-like branches \cite{kkc88pla} on the
current-voltage  characteristics (IVCs) of the LJJ. Studies of these
IVCs for the different shapes of the inhomogeneity in the
genuinely 2D case are in progress and will be published elsewhere.

\section*{Acknowledgemets}

Y.Z. acknowledges financial support from
Ukrainian State Grant for Fundamental Research No.~ 0112U000056.



\end{document}